\documentclass[aps,prl,twocolumn,groupedaddress]{revtex4-1}

\usepackage{graphicx,color}
\usepackage{epsfig}
\usepackage{dcolumn}
\usepackage{amsmath,amssymb,bm}
\usepackage{hyperref,url}
\usepackage[noabbrev]{cleveref}
\usepackage{CJK}
\usepackage{lineno}
\usepackage[english]{babel}
\usepackage[T1]{fontenc}
\usepackage{pifont}
\usepackage{amsfonts}
\usepackage{wasysym}
\usepackage{amsbsy}
\usepackage{psfrag}
\usepackage{color}
\usepackage{multirow}
\usepackage{cases}
\usepackage{stackengine}
\usepackage{float}
\usepackage{blkarray}
\usepackage{cancel}
\usepackage{mathrsfs}
\usepackage{empheq}
\usepackage{mathptmx} 

\DeclareMathAlphabet{\mathbfsf}{\encodingdefault}{\sfdefault}{bx}{sl}

\newcommand{\V}{\mathcal{V}}
\newcommand{\Surf}{\partial \V}

\newcommand{\vzero}{\boldsymbol{0}}

\newcommand{\vx}{\boldsymbol{x}}
\newcommand{\vu}{\boldsymbol{u}}

\providecommand\bnabla{\boldsymbol{\nabla}}
\providecommand\bcdot{\boldsymbol{\cdot}}

\newcommand{\boldm}[1]{\boldsymbol{#1}}

\newcommand{\bnablas}{\boldsymbol{\nabla}_{\! s}}

\newcommand{\Rey}{\text{\textit{Re}}} 	    
\newcommand{\Bo}{\mbox{\textit{Bo}}} 		
\newcommand{\Lap}{\mbox{\textit{La}}} 		
\newcommand{\Bou}{\mbox{\textit{Bq}}} 	    

\hypersetup{
    colorlinks=true,
    linkcolor=blue,
    citecolor=blue,
    urlcolor=blue,
}

\usepackage{fancyhdr}
\fancypagestyle{titlepage}{
\fancyhead[C]{Published in Physical Review Letters \textbf{125,} 114502 (2020)}
\fancyhead[L]{}
}

\begin{document}



\title{Universal Thinning of Liquid Filaments under Dominant Surface Forces}



\author{A. Mart\'inez-Calvo}
\affiliation{Grupo de Mec\'anica de Fluidos, Departamento de Ingenier\'ia T\'ermica
y de Fluidos, Universidad Carlos III de Madrid. Avda. de la Universidad 30, 28911,
Legan\'es, Madrid, Spain.}

\author{A. Sevilla}
\altaffiliation{Corresponding author: alejandro.sevilla@uc3m.es}
\affiliation{Grupo de Mec\'anica de Fluidos, Departamento de Ingenier\'ia T\'ermica
y de Fluidos, Universidad Carlos III de Madrid. Avda. de la Universidad 30, 28911,
Legan\'es, Madrid, Spain.}


\begin{abstract}
Theory and numerical simulations of the thinning of liquid threads at high superficial concentration of surfactants suggest the existence of an asymptotic regime where surface tension balances surface viscous stresses, leading to an exponential thinning with an $e$-fold time $F(\Theta)\,(3\mu_s + \kappa_s)/\sigma$, where $\mu_s$ and $\kappa_s$ are the surface shear and dilatational viscosity coefficients, $\sigma$ is the interfacial tension, $\Theta=\kappa_s/\mu_s$, and $F(\Theta)$ is a universal function. The potential use of this phenomenon to measure the surface viscosity coefficients is discussed.
\end{abstract}

\date{\today}


\maketitle


\emph{Introduction}.-- The adsorption of surfactants at fluid interfaces induces substantial modifications of their mechanical properties that lead to a number of effects of relevance in many physiological and technological contexts~\cite*{VanGolde88,*TweedyARFM,*Rodriguez2015,*Hermans2015,*Anna16}. The variety and complexity of the interactions between the bulk fluids and the surface layer at the microscopic level complicates the development of rigorous mean-field models, that are necessary to describe interfacial dynamics using continuum theories~\citep{FullerVermant2012,Jaensson2018}. The simplest constitutive equation relating the surface stress with the surface rate of strain is the Boussinesq--Scriven (BS) law~\citep{Boussinesq1913,Scriven60}, which may be seen as the surface analog of the Navier--Poisson law. Indeed, the BS law assumes that the surface state of stress is isotropic, instantaneous and linear in the surface rate of strain, and disregards complex surface rheology~\citep{Edwards1991,FullerVermant2012,LangevinARFM,Jaensson2018}, leading to the concept of a \emph{Newtonian surface}~\citep{Scriven60}. The BS law introduces three material parameters, namely the surface tension coefficient, $\sigma(\Gamma,T)$, and the surface shear and dilatational viscosity coefficients~\citep{Rayleigh1890SurfVisc}, $\mu_s(\Gamma,T)$ and $\kappa_s(\Gamma,T)$, respectively, which depend on the surface concentration of surfactant, $\Gamma$, and on the temperature $T$. An important difficulty in the practical use of the BS law concerns the fact that most surfactants are soluble in the bulk fluids, which implies the need to account for bulk diffusion and adsorption-desorption kinetics in the description. In fact, since $\Gamma$ is extremely difficult to measure directly, the bulk concentration of surfactants, $c$, is normally used instead as the experimental control parameter. However, the relationship between $\Gamma$ and $c$ is not universal, but depends on the particular system under study in a nontrivial way that is usually rationalized in terms of appropriate adsorption isotherms~\citep{Prosser2001}. The latter difficulty can be avoided by using high bulk concentrations, typically several times the critical micelle concentration (CMC), in which case the surface concentration is limited by maximum packing and is said to be \emph{saturated}, $\Gamma=\Gamma_{\text{sat}}$, and the corresponding values of the surface tension and surface viscosity coefficients reach corresponding asymptotes $\sigma^{\text{sat}}=\sigma(\Gamma_{\text{sat}},T)$, $\mu_s^{\text{sat}}=\mu_s(\Gamma_{\text{sat}},T)$ and $\kappa_s^{\text{sat}}=\kappa_s(\Gamma_{\text{sat}},T)$~\citep{Edwards1991,Quere1998,Scheid10,Scheid2012}. Moreover, not only surfactants can confer surface viscous resistance to fluid interfaces. Surface viscous forces also arise, for instance in vesicles, biological membranes, or active interfaces, where they coexist with the intrinsic elastic forces, and they may have a dominant role in their dynamics as detailed in Refs.~\citep{Powers2010,Narsimhan2015,Mietke2019,Mietke2019PRL,Farutin2019} and references therein, polymersomes being the prominent example.

For isothermal saturated interfaces, the Marangoni stress is negligible, and the only relevant surface stresses are the Young--Laplace pressure and the surface viscous stresses~\citep{Quere1998,Scheid10,Scheid2012}. Since the surface viscosity coefficients are very difficult to measure under the presence of significant Marangoni stresses and sorption kinetics in that all these effects are intrinsically entangled~\citep{Prosser2001,Elfring2016}, working at saturated conditions opens promising avenues to develop novel measurement techniques. Another difficulty that must be circumvented is the fact that the known values of the surface viscosity coefficients are very small, and thus the corresponding stresses tend to be hindered by bulk stresses. At small Reynolds numbers, the relative importance of the surface-to-bulk viscous stresses is given by the Boussinesq numbers $\Bou=\mu_s/(\mu\ell)$ and $\Theta\,\Bou$, where $\Theta=\kappa_s/\mu_s$ is the dilatational-to-shear surface viscosity ratio, $\mu$ is the viscosity of the bulk fluid, and $\ell$ is the characteristic length scale. Consequently, for the surface viscous stresses to be larger than the bulk viscous stresses it is necessary that $\Bou > 1\Rightarrow \ell < \mu_s/\mu$, where the length scale $\mu_s/\mu$ plays here a similar role as the \emph{Saffman-Delbr\"uck length} in membrane biophysics~\citep{Saffman75,saffman76,stone1995fluid,Powers2010}. As outlined by Ref.~\citep{Powers2010}, this length can be $\ell \approx 1 \mu$m for liposomes~\citep{Dimova1999,Dimova2000,Dimova2006}, and $\ell \approx 1$ mm for polymersomes~\citep{Dimova2002}, where surface viscous forces dominate. For example, diblock copolymer vesicles have associated surface viscosities which are up to 500 times higher than the typical values. These small scales can be reached by means of several thinning mechanisms, e.g. the drainage of foams and emulsions~\citep{Joye94,Hermans2015}, the formation and drainage of thin films~\citep{Scheid10,Champougny2015,Bhamla2017,Seiwert2014,Ozan2019}, the lifetimes of antibubbles or bubbles bursting at a free surface~\citep{Dorbolo2005,Scheid2012,Vitry2019}, or the dynamical necking processes leading to the pinch-off of liquid bridges~\citep{Liao2006,Ponce16b,Kovalchuk2018}, dripping faucets~\citep{Ponce17}, or vesicles and biological membranes~\citep{Bar1994,Powers2010,Narsimhan2015,Ruiz2019,Tozzi2019,Mietke2019,Mietke2019PRL,Farutin2019}.

\begin{figure*}[ht!]
  \begin{center}
    \includegraphics[width=\textwidth]{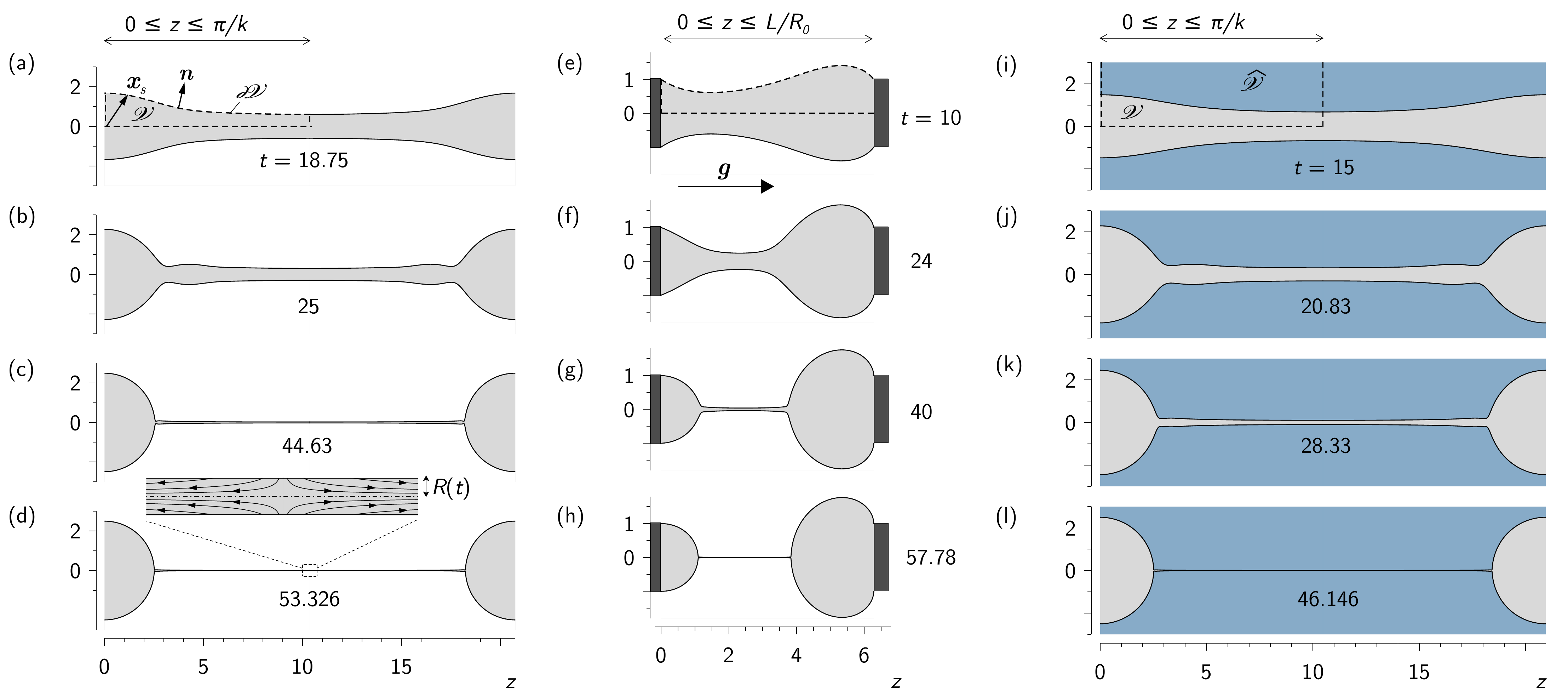}
    \caption{\label{fig:figure1}Snapshots of the liquid thread extracted from the numerical simulations. (a)-(d) Case I with $\Bou = 3.4$, $\Theta = 1$ and $k = 0.3$. (e)-(h) Case II with $\Bou = 0.1$, $\Theta = 2$, $L/R_0=2\pi$ and $\Bo=0.3$. (i)-(l) Case III with $\Bou = 2$, $\Theta = 3$, $N_{\mu} = 10^{-3}$ and $k=0.3$.}
  \end{center}
\end{figure*}

In this Letter, we report theoretical and numerical evidence pointing to a new asymptotic regime where surface tension balances surface viscous stresses. To that end, we consider the small scales generated by the thinning of axisymmetric liquid filaments due to capillary drainage, whereby the dominant force balance is $\sigma/R^2 \sim \mu_s \dot{R}/R^3$, where $R(t)$ is the filament radius and $\dot{R}={\rm d}R/{\rm d}t$ its associated radial velocity. The latter balance assumes that $R\ll \mu_s/\mu$, i.e. that the local Boussinesq number $\Bou_{\ell}=\mu_s/(\mu\,R)\gg 1$, and anticipates the existence of the exponential thinning regime $R(t) \propto \exp{(-t/t_c)}$, where $t_c \sim \mu_s/\sigma$ is a characteristic time that depends only on material parameters~\citep{MartinezSevilla2018}. Note that $t_c$ may be called the surface-viscocapillary time in analogy with the classical viscocapillary time, $\mu \ell/\sigma$, given by the balance of surface tension and bulk viscous forces, that, in contrast with its surface analogue, depends on the length scale $\ell$. In the absence of surfactants, the local balance of surface tension and bulk viscous forces, $\sigma/R^2 \sim \mu \dot{R}/R^2$, provides $\dot{R}\sim \sigma/\mu$, i.e., a thinning at the viscocapillary velocity $\sigma/\mu$~\citep{Rayleigh4,Papageorgiou1995}.


\emph{Numerical simulations}.-- We performed numerical simulations of the Stokes equations for three different axisymmetric flow configurations (see Fig.~\ref{fig:figure1} and the movies provided as Supplemental Material). In case I, we studied the spatially periodic dynamics of a long viscous liquid thread inside an unbounded passive ambient in the absence of gravity. The liquid filament was destabilized by a small-amplitude harmonic disturbance of the cylindrical shape with a wavenumber $k$ below the Plateau--Rayleigh cut-off~\cite{Ashgriz1995,Kamat2018,MartinezSevilla2018,martinez2020natural}. In case II, we considered the unstable dynamics of a liquid bridge between two solid cylinders surrounded by a passive ambient, with gravity pointing in the axial direction and an associated Bond number $\Bo=\rho g R_0^2/\sigma$, where $\rho$ is the liquid density. The bridge length is fixed above the critical length for spontaneous breakup due to the Plateau--Rayleigh instability~\cite{Liao2006,Ponce16b,Kovalchuk2018}. Finally, case III was the same as case I, but with the liquid filament surrounded by an immiscible ambient liquid~\cite{Tomotika,Hajiloo87,StoneLeal1990,Tjahjadi1992,Milliken1993,stone1996note,Gaudet1996,Hansen99}.

\begin{figure*}[ht!]
  \begin{center}
    \includegraphics[width=\textwidth]{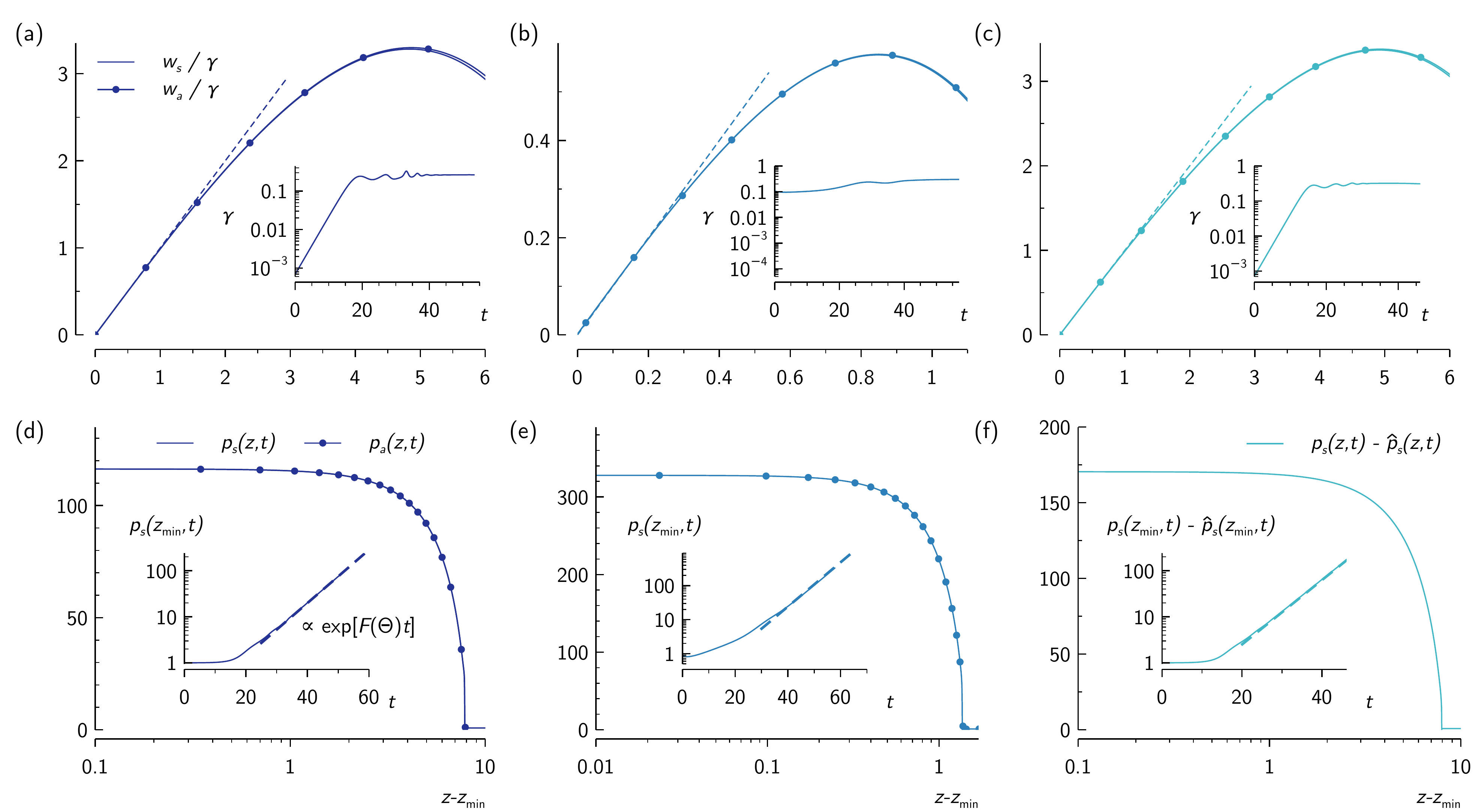}
    \caption{\label{fig:figure2}(a)-(c) Axial velocity normalized with the local strain rate evaluated at the axis, $w_a/\gamma$, and at the interface, $w_s/\gamma$, as functions of $z$ for (a) case I at $t = 53.326$, (b) case II at $t = 57.780$, and (c) case III at $t = 46.146$ [snapshots (d), (h) and (l) in Fig.~\ref{fig:figure1}]. The insets display the local strain rates, $\gamma(t)$. (d),(e) Axial profiles of the pressure at the axis, $p_a(z,t)$, and at the interface, $p_s(z,t)$, for cases I and II. (f) Axial profile of the pressure jump at the interface, $p_s(z,t)-\hat{p}_s(z,t)$, for case III. The insets in (d)-(f) show $p_s(z = z_{\min},t)$ (d),(e), and $p_s(z_{\min},t)-\hat{p}_s(z_{\min},t)$ (f), where $z_{\min}$ is the axial position of minimum radius, with $z_{\min}=\pi/k$ in cases I and III.}
  \end{center}
\end{figure*}

Taking the initial thread radius, $R_0$, as the length scale, the surface-viscocapillary time, $(3\mu_s+\kappa_s)/\sigma$, and velocity, $\sigma\,R_0/(3\mu_s+\kappa_s)$, as the time and velocity scales, and the capillary pressure, $\sigma/R_0$, as the pressure scale, the dimensionless Stokes equations read:
\begin{subequations}\label{eq:continuity_momentum}
\begin{gather}
\bnabla \bcdot \boldm{u} = 0, \quad \text{and} \quad \boldm{0} = \bnabla \bcdot \mathbfsf{T}
 \quad \text{in} \quad \V, \label{eq:inner}\\
\bnabla \bcdot \hat{\boldm{u}} = 0, \quad \text{and} \quad \boldm{0} = \bnabla \bcdot \hat{\mathbfsf{T}} 
 \quad \text{in} \quad \hat{\V},\label{eq:outer}
 \end{gather}
\end{subequations}
where hatted quantities correspond to the outer fluid, $\boldm{u}(\vx,t) = u \, \boldm{e}_r + w \, \boldm{e}_z$ is the velocity field, $u$ and $w$ being the radial and axial velocity components, and $\mathcal{V}$, $ \hat{\V}$ are the inner and outer fluid domains. The bulk stress tensors are:
\begin{subequations}\label{eq:bulk_stress_tensors}
\begin{gather}
\mathbfsf{T} = - p \mathbfsf{I} + \frac{1}{\Bou (\Theta + 3)}\,[\bnabla \boldm{u} + (\bnabla \boldm{u})^{\rm{T}}], \label{eq:bulk_stress_tensor_inner}\\
\hat{\mathbfsf{T}} = -\hat{p}\mathbfsf{I} + \frac{N_{\mu}}{\Bou (\Theta + 3)}\,[\bnabla \hat{\boldm{u}} + (\bnabla \hat{\boldm{u}})^{\rm{T}}],\label{eq:bulk_stress_tensor_inner}
\end{gather}
\end{subequations}
where $\mathbfsf{I}$ is the identity tensor, $p(\vx,t)$ is the pressure field, $\Bou=\mu_s/(\mu R_0)$ is the Boussinesq number, and $N_{\mu} = \hat{\mu}/\mu$ is the outer-to-inner viscosity ratio. Note that the relative importance of liquid inertia compared with the viscous forces is measured by a local Reynolds number $\Rey_{\ell}=\Lap R |\dot{R}|$, where $\Lap = \rho \sigma R_0/\mu^2$ is the Laplace number. Anticipating that $R \sim \exp{(-t)}$ for $t\gg 1$, it is deduced that $\Rey_{\ell} \sim \Lap \exp{(-2t)}$, indicating that inertia becomes negligible at large times. The interfacial stress balance reads:
\begin{equation}\label{eq:interf_stress}
(\hat{\mathbfsf{T}} - \mathbfsf{T}) \bcdot \boldm{n} + \bnablas \bcdot \mathbfsf{T}_s = \boldm{0} \quad \text{at} \quad \Surf,
\end{equation}
where $\boldm{n}$ is the outer normal to the interface and $\mathbfsf{T}_s$ is the surface stress tensor, modeled using the BS law~\citep{Boussinesq1913,Scriven60},
\begin{align}\label{eq:boussinesqtensor}
& \mathbfsf{T}_s = \left[1 +  \frac{\Theta - 1}{\Theta+3}(\bnablas \bcdot \vu_s) \right]\mathbfsf{I}_s + & \nonumber \\
&  \frac{1}{\Theta+3} \left[(\bnablas \vu_s)\bcdot \mathbfsf{I}_s+\mathbfsf{I}_s\bcdot (\bnablas \vu_s)^{\text{T}}\right],
\end{align}
where $\boldm{u}_s$ is the fluid velocity at the interface, $\mathbfsf{I}_s = \mathbfsf{I}-\boldm{n} \boldm{n}$ is the surface projection tensor, and $\bnablas=\mathbfsf{I}_s\bcdot \bnabla$. As explained in the introduction, $\mu_s$ and $\kappa_s$ generally depend on the surface concentration of surfactant~\citep{Edwards1991}, unless the region adjacent to the interface is highly populated with surfactant molecules, and their adsorption time is much smaller than the characteristic hydrodynamic time. In the latter saturated limit, the effect of the Marangoni stress $\bnablas\sigma$ becomes negligible~\citep{Quere1998,Scheid10,Scheid2012}, and thus a surfactant transport equation is not needed to close the mathematical model. The BS law has also been used in the context of passive and active vesicles and membranes~\citep{Powers2010,Narsimhan2015,Farutin2019,Mietke2019,Mietke2019PRL}, in the latter case coupled with continuum theories borrowed from nematic and active-gel theories~\citep{Marchetti2013,Prost2015,Hakim2017,Julicher2018,Farutin2019,Mietke2019,Mietke2019PRL}. Additionally, at the interface $\partial \mathcal{V}$ we impose the continuity of velocities, $\hat{\boldm{u}} = \boldm{u}$, and the kinematic condition, $\boldm{u}_s  \bcdot \boldm{n}   = \dot{ \boldm{x}}_s \bcdot \boldm{n}$,
where $\boldm{x}_s$ is the parametrization of the interface [see Fig.~\ref{fig:figure1}(a)]. The boundary conditions in the $z$ direction are $\partial_z u = w = 0$ at $z=0,\pi/k$ for cases I and III, while in case II we impose $u = w = 0$ at $z=0, L/R_0$, where $L$ is the length of the liquid bridge. In the three cases, the axisymmetry condition $\partial_r w = u = 0$ holds at the axis, $r=0$.
As for the initial conditions, we assume that the fluids are initially at rest, $\vu = \vzero$, and impose shape disturbances of the form $\vx_s= [1 + \epsilon \cos (k z)] \boldm{e}_r + z \, \boldm{e}_z$ in cases I and III, and $\vx_s= \left\{1+\epsilon\left[\cos\left(2 \pi z/(L/R_0) \right)-1\right] \right\} \boldm{e}_r + z \boldm{e}_z$ in case II, where $\epsilon \ll 1$ is a small disturbance amplitude. 
We now have a closed system to determine $\boldm{u}$, $p$ and $\vx_s$ in cases I and II, and additionally $\hat{p}$ and $\hat{\boldm{u}}$ in case III. The numerical integration employs the same methodology explained in previous studies~\citep{RiveroScheid2018,martinez2020natural}, where a detailed description can be found.


\begin{figure*}
  \begin{center}
    \includegraphics[width=\textwidth]{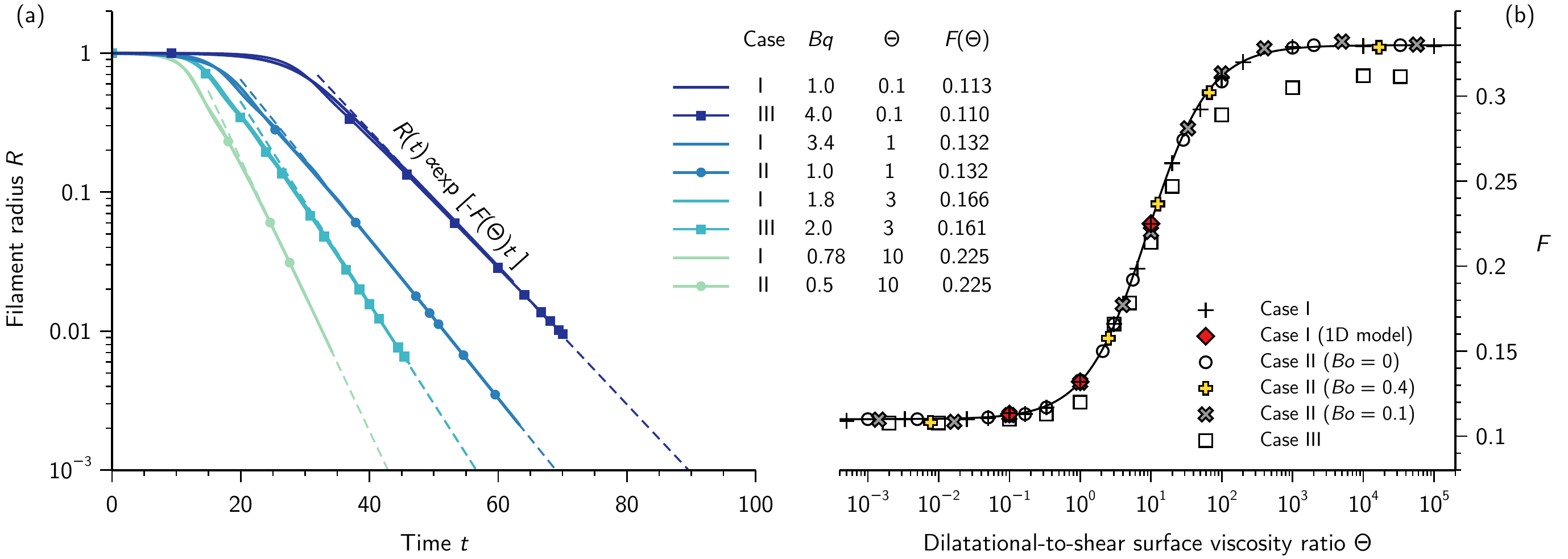}
    \caption{\label{fig:figure3}(a) The minimum thread radius $R(t)$ for different conditions specified in the legend. (b) The universal function $F(\Theta)$. The solid line corresponds with the approximation $F(\Theta) \approx (3+\Theta)/(28+3.02 \Theta)$.}
  \end{center}
\end{figure*}


\emph{Numerical results}.-- Figure~\ref{fig:figure1}, and the corresponding movies provided as Supplemental Material, show representative interface evolutions for the three cases under study. The computational domains are indicated in the snapshots (a),(e),(i) with dashed lines, and the inset in (d) shows the local flow near the symmetry plane of the elongated thin thread connecting the two main drops. As anticipated before, the surface viscous stresses avoid the occurrence of the finite-time singularity that would lead to pinch-off if only bulk viscous stresses balanced interfacial tension~\citep{Papageorgiou1995}. Instead, the radius of the cylindrical filament is observed to relax exponentially at large times, as evidenced in Fig.~\ref{fig:figure3}(a), which also shows that the thread relaxation rate increases monotonically with $\Theta$. It is noteworthy that these exponentially decaying filamentary structures resemble the celebrated beads-on-a-string structure in viscoelastic liquid threads~\citep{Goldin1969, Bazilevskii1981, Entov1997, Clasen2006}. Nevertheless, the physical mechanisms underlying both phenomena are completely different, as demonstrated below.


\emph{Local analysis of the large-time behavior}.-- To develop a simple theory that accounts for the exponential relaxation of the liquid thread, we examined the numerical evidence carefully. In particular, motivated by the shape evolution shown in Fig.~\ref{fig:figure1}, we approximate the thinning ligament by a cylinder of radius $R(t)$, and we assume that the axial velocity inside the ligament is uniform in the radial direction, $w=w(z,t)$, as evidenced by the profiles of axial velocity extracted at the axis, $w_a(z,t)$, and at the interface, $w_s(z,t)\approx w_a(z,t)$, represented in Figs.~\ref{fig:figure2}(a)-\ref{fig:figure2}(c). For simplicity, we decided to develop the local model disregarding the bulk viscous stresses of the outer flow, so that $\hat{\mathbfsf{T}}\approx -\hat{p}\mathbfsf{I}$. Indeed, although a cylindrical interface cannot be an exact solution of~\eqref{eq:continuity_momentum} when $N_{\mu}\neq 0$, we will show below that the exponential thinning regime occurs when the inner and outer bulk stresses are both negligible compared to the surface stresses. The continuity equation in~\eqref{eq:inner} implies that the radial velocity $u(r,z,t)=-\gamma r/2$, where $\gamma=\partial_zw$ is the axial strain rate. Moreover, the kinematic condition applied at $r=R(t)$ implies that $\gamma(t)=-2\dot{R}(t)/R(t)$ is only a function of time, and thus $w(z,t)=\gamma(t)z$, as observed in Figs.~\ref{fig:figure2}(a)-\ref{fig:figure2}(c) in the region $0\leq z - z_{\min}\lesssim 2$. It is thereby deduced that the local elongational flow field $\boldm{u}(r,z,t) = -2z\dot{R}/R\,\boldm{e}_z + r\dot{R}/R\,\boldm{e}_r$
provides a good description of the local dynamics inside the filament. According to the Stokes equation for the inner stream~\eqref{eq:inner}, the latter velocity field is an exact solution provided that the pressure field depends only on time, $p = p(t)$, in agreement with the results of Figs.~\ref{fig:figure2}(d)-\ref{fig:figure2}(f), which show that the pressure field inside the thread is approximately uniform in the region where $w=\gamma(t)z$. The dynamics of the thread is then given by the function $R(t)$, which is determined from the interfacial stress balance~\eqref{eq:interf_stress}. In particular, the surface stress tensor~\eqref{eq:boussinesqtensor} simplifies to $\mathsf{T}_{s}^{zz} = 1- \dot{R}/R, \quad  \text{and} \quad \mathsf{T}_{s}^{\theta \theta} = 1 +  [(3 -\Theta)/(3+ \Theta)] \dot{R}/R$,
the remaining entries being null, whereby the surface stress and the bulk stress jump at the interface read, respectively,
\begin{subequations}\label{eq:tractions}
\begin{gather}
\bnablas \bcdot \mathbfsf{T}_{s} = -\left(\frac{1}{R} + \frac{3 - \Theta}{3 + \Theta}\frac{\dot{R}}{R^2} \right) \boldm{e}_r + \partial_z \left( 1- \frac{\dot{R}}{R}\right) \boldm{e}_z,\label{eq:surf_traction}\\
\left(\hat{\mathbfsf{T}} - \mathbfsf{T} \right) \bcdot \boldm{n} = \left(p_s - \hat{p}_s - \frac{2}{\Bou (3+\Theta)} \frac{\dot{R}}{R}\right) \boldm{e}_r + \frac{\partial_z  \dot{R}}{\Bou (3+\Theta)}\,\boldm{e}_z.\label{eq:bulk_traction}
\end{gather}
\end{subequations}
Using Eqs.~\eqref{eq:interf_stress} and~\eqref{eq:tractions} we obtain the normal and tangential interfacial stress balances. For $\Theta\neq 3$ and $t \gg 1$, the normal component reduces to $\dot{R} = - F(\Theta)\,R$,
where we have defined the positive-definite function $F(\Theta)=[(3+\Theta)/(3-\Theta)]\,\left[1-\lim_{t\to\infty} (p_s-\hat{p}_s)R\right]$, represented for the three cases in Fig.~\ref{fig:figure3}(b) together with the approximation $F(\Theta) \approx (3+\Theta)/(28+3.02 \Theta)$, which provides a good fit to the numerical results. Note that the value $F(\Theta=3)\simeq 0.166$ is obtained using the tangential stress balance. Finally, the thread radius obeys $R(t)\propto \exp\left[-F(\Theta)\,t\right]$, in close agreement with all the numerical results represented in Fig.~\ref{fig:figure3}(a).


\emph{Discussion and applicability conditions}.-- The assumptions of constant surface viscosities and negligible Marangoni stress clearly need some justification. As discussed by~\citet{Quere1998} and by~\citet{Scheid10} in the context of axisymmetric and planar coating flows, respectively, two conditions must be fulfilled to ensure the validity of these hypotheses. First, as the thread shrinks, there must be enough available surfactant at the sublayer adjacent to the interface. This condition is satisfied when $c\gg \Gamma_{\text{sat}}\,R_0^{-1}$, and can be easily guaranteed in experiments by means of a liquid bath with a high concentration of surfactant surrounding the inner thread~\footnote{B. Scheid (private communication)}, as in the simulations of case III considered herein. Moreover, it could be advantageous to use outer liquids of small viscosity, $N_{\mu}\ll 1$, to replenish the interface with surfactants, in that the ambient fluid acts passively, as in cases I and II of the present investigation. Second, the adsorption velocity must be larger than the characteristic interfacial velocity, what implies that the flux from the bulk, $k_a\,c$, where $k_a$ is the adsorption velocity, is much larger than the surface flux $\Gamma_{\text{sat}}\,\sigma^{\text{sat}}(3\mu_s^{\text{sat}}+\kappa_s^{\text{sat}})^{-1}\,\dot{R}R^{-1}$, requiring that $c\gg \Gamma_{\text{sat}}\,k_a^{-1}\,\sigma^{\text{sat}}(3\mu_s^{\text{sat}}+\kappa_s^{\text{sat}})^{-1}\,F(\Theta)$, a condition that, again, can be accomplished using a highly concentrated outer bath. Finally, it is interesting to note that the results obtained herein using a fully two-dimensional Stokes and Boussinesq-Scriven description cannot be deduced from the one-dimensional lubrication approximation derived in Ref.~\citep{MartinezSevilla2018}
which, assuming that the curvature is $h^{-1}$ in the surface viscous terms, has the conservation form
\begin{equation}\label{eq:1D_conserv}
0 =  \partial_z \left[h^2 \mathcal{K} + \frac{3 \, h^2 \partial_z u}{\Bou (\Theta+3)}+\frac{ h \partial_z u  (\Theta+9)}{2(\Theta+3)} \right],
\end{equation}
where $\mathcal{K} = h^{-1} (1+h'^2)^{-1/2}+h''(1+h'^2)^{-3/2}$~\citep{Entov1984,Clasen2006} and primes indicate derivatives with respect to $z$, together with the continuity equation $\partial_t h^2 + \partial_z (h^2 u) = 0$. Integrating Eq.~\eqref{eq:1D_conserv} yields a function of time $\lambda(t)$ that can be seen as the total force acting on the filament~\cite*{Li2003,*Fontelos2004}, with $\lambda(t) \sim R(t)$ to balance the capillary pressure term, so that $\lambda(t)/R(t)\to \Lambda$ for $t \gg 1$, where $\Lambda$ is a function of $\Theta$ only. Although purely cylindrical solutions of Eq.~\eqref{eq:1D_conserv} obey an equation similar to that deduced above from the Stokes equations, the parameter $\Lambda$ can only be related with the liquid pressure if the full Eqs.~\eqref{eq:continuity_momentum}--\eqref{eq:boussinesqtensor} are considered, providing $\lim_{t \to \infty} (p_s-\hat{p}_s)R = 1-(3-\Theta)(\Lambda -1)/(9+\Theta)$, as deduced also with the second-order parabolic model~\cite{MartinezSevilla2018}. 


\emph{Concluding remarks}.-- We have shown that a fluid interface saturated with surfactant molecules displays an exponential capillary thinning regime where surface viscous stresses balance surface tension. This new dynamical regime had not been reported in previous numerical~\citep{Ponce16b,Ponce17} and experimental~\citep{Liao2006,Ponce17,Kamat2018,Kovalchuk2018} investigations probably because, in all these studies, the depletion of surfactant due to interfacial advection was not compensated by an outer reservoir able to replenish the interface. Our findings could well open promising avenues in developing novel techniques for the high-precision measurement of the surface viscosity coefficients~\citep{Jaensson2018}.


\begin{acknowledgments}
We are grateful to B. Scheid and J. Rivero-Rodr\'iguez for useful comments and numerical advice. A.M.-C. also acknowledges fruitful and inspiring discussions with A. Perazzo during his stay in Princeton, and support from the Spanish MECD through the Grant No. FPU16/0256. This research was funded by the Spanish MCIU-Agencia Estatal de Investigaci\'on through project DPI2017-88201-C3-3-R, partly financed through FEDER European funds. Support from the Red Nacional para el Desarrollo de la Microflu\'idica, RED2018-102829-T, is also acknowledged.
\end{acknowledgments}

\vspace{0.3cm}

\emph{Note added}.-- We recently became aware of a similar study reporting the exponential thinning of liquid threads in the limit of dominant surface diffusion~\citep{Wee2020}.

\bibliography{biblio}

\end{document}